\documentclass{emulateapj} 
\usepackage{enumerate,multirow}
\usepackage{amsmath,amssymb,graphicx,natbib,subfigure} 

\usepackage{amsmath}
\usepackage{amssymb}
\usepackage{color}

\shorttitle{The imprint of black-hole mergers on nuclear star clusters}
\shortauthors{Antonini, Barausse and Silk}

\begin{document}
\def\gap{\;\rlap{\lower 2.5pt
\hbox{$\sim$}}\raise 1.5pt\hbox{$>$}\;}
\def\lap{\;\rlap{\lower 2.5pt
 \hbox{$\sim$}}\raise 1.5pt\hbox{$<$}\;}

\def\eb#1{{\textcolor{blue}{[[\bf EB: #1]]}}}
\def\mb#1{{\textcolor{red}{[[\bf MB: #1]]}}}

\newcommand\MBH{\rm MBH} 
\newcommand\NC{NSC}
\newcommand\CMO{CMO}

\title{The imprint of massive black-hole mergers
on the correlation between nuclear star clusters and their host galaxies}

\author{Fabio Antonini$^{1}$, Enrico Barausse$^{2,3}$ \& Joseph Silk$^{2,3,4,5}$}
\affil{(1) Center for Interdisciplinary Exploration and Research in Astrophysics (CIERA)
and Department of Physics and Astrophysics, 
Northwestern University,  Evanston, IL 60208\\
(2) Sorbonne Universit\'{e}s, UPMC Univ Paris 06, UMR 7095, Institut d'Astrophysique de Paris, F-75014, Paris, France\\
(3) CNRS, UMR 7095, Institut d'Astrophysique de Paris, F-75014, Paris, France \\
(4) Laboratoire AIM-Paris-Saclay, CEA/DSM/IRFU, CNRS, Universite Paris Diderot,  F-91191 Gif-sur-Yvette, France   \\
(5) Department of Physics and Astronomy, Johns Hopkins University,  Baltimore MD 21218, USA
}

\begin{abstract}
A literature compilation of nuclear star cluster (\NC) masses is used 
to study the correlation between global and nuclear  properties. 
A comparison of observational data 
to the predictions of semi-analytical galaxy formation
models places constraints on the co-evolution of \NC s, massive black holes~(\MBH s)
and host galaxies. 
Both data and theoretical predictions show an increased scatter in the 
\NC \ scaling correlations at high galaxy masses, and we 
show that this is due to
  the progressively more efficient 
 ejection of stars from \NC s caused by \MBH \ binaries in more massive 
stellar spheroids. 
Our results  provide a natural explanation of why 
in nucleated galaxies hosting a \MBH,
the ratio  $(M_{\rm \NC}+M_{\rm \MBH})/M_{\rm bulge}$ (with 
$M_{\rm bulge}$  the host spheroid's mass) shows significantly less scatter than 
 $M_{\rm \NC}/M_{\rm bulge}$, and suggest that  the formation of \MBH s and \NC s are not
mutually exclusive, as also supported by observations of co-existing systems. 
Both \MBH s and \NC s represent generic products of galaxy formation, 
with \NC s being destroyed or modified by the merger evolution of their companion \MBH s.
\end{abstract}
 

\section{Introduction}
Observations over the last two decades have
unveiled the existence 
of compact stellar nuclei  at the centers of most intermediate and 
low luminosity 
galaxies~\citep{1998AJ....116...68C,2002AJ....123.1389B,2014MNRAS.445.2385D,2014MNRAS.441.3570G}. 
With masses of order $0.1\% $
the stellar mass of the host galaxy and typical radii of  only a few parsecs, these clusters,
often referred to as  \NC s, are the
densest stellar systems  observed in the 
Universe~\citep[e.g.,][]{2005ApJ...618..237W,Cote:2006,2015arXiv150105586C}. 

The masses of \NC s  ($M_{\rm \NC}$) obey fairly  tight correlations 
with properties  of the host galaxies such as the
bulge velocity dispersion ($\sigma$) and mass ($M_{\rm bulge}$)~\citep{Ferrarese:2006}.
A number of authors have shown that these correlations are much shallower than the corresponding ones 
for \MBH s.   \citet{G12} finds that \NC \ masses obey
$M_{\rm \NC} \sim \sigma^2$, while the masses of  \MBH s, $M_{\rm MBH}$, follow a much steeper scaling relation
 $M_{\rm MBH}\sim \sigma^5$~\citep{2005SSRv..116..523F}. 
It is intriguing that for galaxies containing both a \NC\ and a \MBH ,
 the ratio 
$M_{\rm \CMO}/M_{\rm bulge}$ (with $M_{\rm \CMO}=M_{\rm MBH}+M_{\rm \NC}$ the 
total mass in central massive objects) shows less scatter than 
 either $M_{\rm MBH}/M_{\rm bulge}$ or $M_{\rm \NC}/M_{\rm bulge}$~\citep{2013ARA&A..51..511K}.
 The facts that both \NC s and \MBH s are found to follow tight correlations with their host galaxy properties,
 that they are found to coexist in some galaxies, and that in these galaxies they have comparable masses,
 point toward a picture where 
\NC s and \MBH s  are both generic by-products of galaxy
formation, and the growth mechanisms of \NC s and \MBH s
are related to one another and to the host galaxy evolution.

We have developed a semi-analytical model 
following the formation and evolution of galaxies, \MBH s and \NC s along cosmic history.
Previous calculations assumed \NC\ formation to take place in isolated 
galaxies~\citep[e.g.,][]{2011ApJ...729...35A,2013ApJ...763...62A,2014ApJ...785...71G,2014MNRAS.444.3738A,2015ApJ...799..185A}, 
thus neglecting the possible role of galaxy mergers, as well as in-situ star formation processes. 
Also, these early idealized attempts did not explore the possible interplay between
\MBH \ and \NC \ evolution. 
Our model sheds light on exactly these points, i.e. it allows us to assess for the first time the role of galaxy mergers, \MBH\ mergers
and nuclear star formation on the formation and evolution of \NC s.

\begin{figure*}
\centering
\includegraphics[width=2.35in,angle=0.]{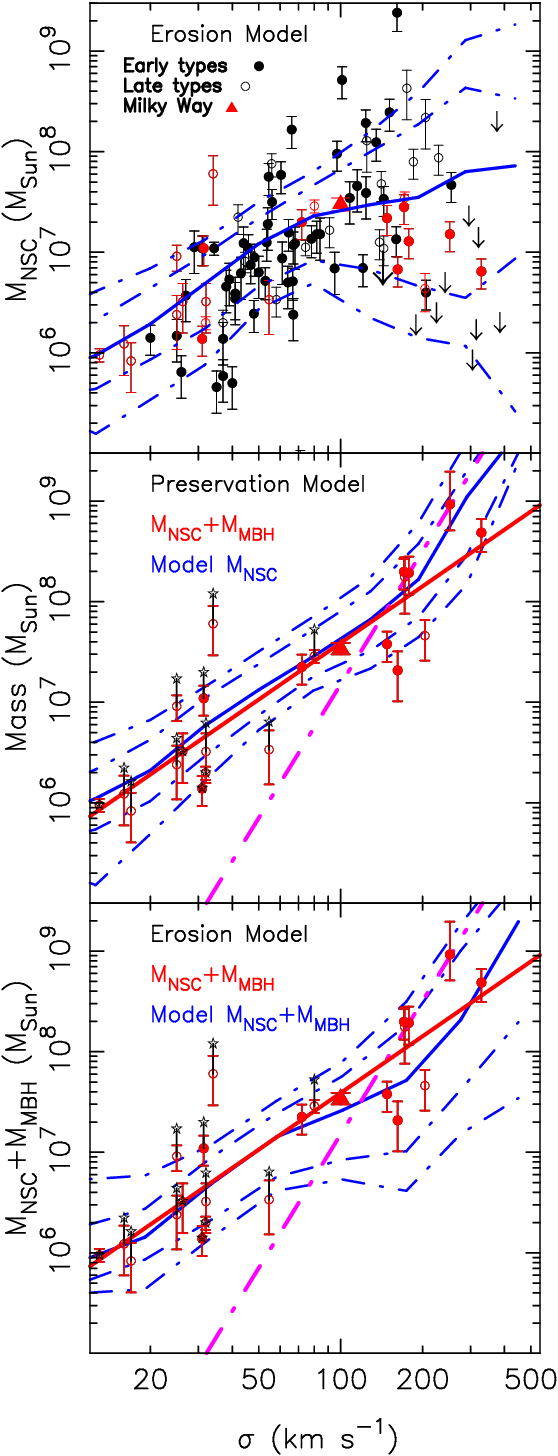}~
\includegraphics[width=2.31in,angle=0.]{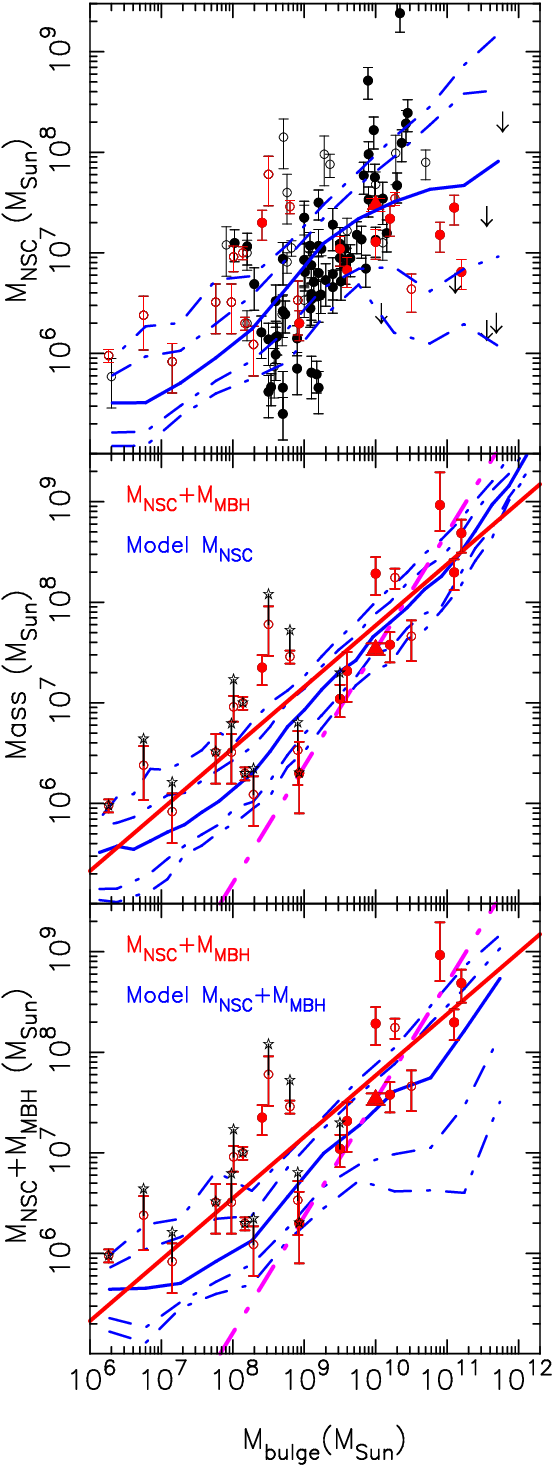}
\caption{Comparison of our model to the observed \NC\ scaling relations.
The upper and lower panels present respectively the predicted 
$M_{\rm \NC}$ and $M_{\rm \NC}+M_{\rm \MBH}$  for the ``erosion'' model,
which accounts for the mass deficit caused by \MBH\ binaries,
while the middle panels give  $M_{\rm \NC}$  for the ``preservation'' model, where this effect is absent.
Blue lines are the median correlations predicted by our model;
 dot-dashed lines represent regions  containing $70\%$ and
$90\%$ of the \NC s produced in our models at a given stellar mass. 
Arrows  are 
upper limits to the mass of \NC s.
Red symbols denote galaxies  with 
 observationally constrained \MBH \ masses; the red symbols  in the different  panels 
 correspond to the same 
 subset of galaxies, but  in the upper panels
they give $M_{\rm \NC}$,  in the bottom and middle panels they give $M_{\rm \NC}+M_{\rm \MBH}$.
For galaxies with only an upper limit to \MBH \ mass, we set $M_{\rm MBH}=0$, but
star symbols show
how  these points would move if we were to assume \MBH \ masses  equal to these upper limits.
 The red lines are fits to the total \NC +\MBH\ masses, while
dot-dashed-purple lines show for comparison the \MBH\ -- host galaxy scaling correlations~\citep{2013ARA&A..51..511K}.
}\label{scaling-r}
\end{figure*}

\section{Method and model}
The backbone of this study is the semi-analytical galaxy formation model of~\citet{2012MNRAS.423.2533B} (with the
improvements described in \citet{2014ApJ...794..104S}).  
This model tracks the evolution of baryonic structures (both in central galaxies and in satellites) 
along dark matter merger trees 
produced with an extended Press-Schechter formalism, modified to reproduce 
the results of $N$-body simulations of dark matter halos~\citep{2008MNRAS.383..557P}. The baryonic structures 
ultimately originate from the hot, largely unprocessed intergalactic medium, which cools to 
form cold gaseous galactic disks. These undergo quiescent star formation, 
which chemically  enriches the  interstellar medium and 
gives rise to stellar disks. Both stellar and gaseous disks are disrupted by bar instabilities --
giving rise to pseudo-bulges that undergo a quiescent disk-like star-formation process --
and by major mergers -- which form classical bulges with violent star-formation bursts,  possibly triggered by turbulent gas flows. Besides
further enriching the interstellar medium, bar instabilities and (mainly) star-bursts 
are assumed to funnel cold, chemically enriched gas into a low-angular momentum
nuclear reservoir, available for accretion onto the central \MBH~\citep{granato,haiman}.
 Possible mechanisms capable of removing angular momentum from the gas are shocks (caused e.g. by mergers) or
radiation drag from star-bursts~\citep{2001ApJ...560L..29U,granato}, and we thus
assume a linear correlation between the feeding of the nuclear reservoir and the star-formation rate
in the bulge (or pseudo-bulge).
The \MBH s may form from either high-redshift seeds
arising from Pop III stars~\citep[i.e. $\sim 100 M_\odot$ ``light'' seeds;][]{2001ApJ...551L..27M} or 
from the remnants of protogalactic disks~\citep[i.e. $\sim 10^5 M_\odot$ ``heavy'' seeds;][]{2004MNRAS.354..292K,2008MNRAS.383.1079V}.
In the former case, we allow moderately super-Eddington accretion onto the massive \MBH s so as to better reproduce
the quasar luminosity function at high redshift~\citep{light_seeds}.

Here, we improve on this model by supplementing it with semi-analytical prescriptions for
the formation and evolution of \NC s. These can grow either by local star formation 
in the nuclear regions~\citep[e.g.,][]{2006AJ....132.2539S},
or by infall of stellar clusters~\citep{1975ApJ...196..407T} toward the center of the galaxy, as a result of dynamical friction. 

In more detail, we assume that the nuclear reservoir from which the \MBH\ accretes \textit{also} undergoes
quiescent disk-like star formation -- for which we adopt the same star-formation
prescription as in galactic disks and pseudobulges, i.e. that of \citet{krum} -- and we then
assume that the stars formed in this nuclear reservoir contribute to the \NC. 
 Note that even though we model star formation in this nuclear gas reservoir as quiescent, our models also assume
that the reservoir's feeding is correlated with star formation in the bulge, and thus mainly with turbulence-driven starbursts following major mergers.

As for the infall of stellar clusters, we assume that the latter form with an efficiency $f_{\rm gc}$ during star-formation episodes in both
disks and bulges/pseudo-bulges, and with initial mass function ${\rm d}n/{\rm d}M\propto M^{-2}$~\citep{bik}, between $M=10^2~M_{\odot}$ and 
$M=10^7~M_{\odot}$. Initially, these stellar clusters follow the same spatial distribution as the stars, but after their formation
we follow their (average) infall towards the \NC \ 
under dynamical friction from both the stars and gas. In doing so, we also account for both tidal heating and evaporation of the stellar clusters
due to the galaxy's tidal field, and for the tidal disruption by the \MBH\ when they approach the center. 
When  the \MBH \ mass becomes $\gtrsim 10^{8-9}~M_{\odot}$, the stellar clusters are
completely disrupted before reaching the center, and the contribution
of stellar cluster infall to \NC \ growth becomes negligible~\citep{2013ApJ...763...62A}.

Whenever two galaxies merge, their \MBH s are assumed to form a binary. 
Due to the interactions with the stars in the nuclear region, a \MBH \ binary ejects from the system a  mass (in stars) comparable to its own mass~\citep{2006ApJ...648..976M}. 
Also, when a \MBH \ binary finally merges, the anisotropic emission of gravitational waves imparts a kick (of up to several 
thousands of km/s, cf. \citet{2007PhRvL..98w1102C}) to the \MBH \ remnant, and this also removes stars from the nuclear region. We thus model
the total mass deficit $M_{\rm ej}$ caused by the inspiral and merger of  \MBH \ binaries onto the \NC \ mass by
\begin{eqnarray}\label{ejmass}
M_{\rm ej}&\approx & 0.7q^{0.2}M_{\rm bin}  \nonumber
+0.5 M_{\rm bin} \ln \left({a_{\rm h}\over a_{\rm gr}}\right) \\
&&+5M_{\rm bin} \left(V_{\rm kick}/V_{\rm esc} \right)^{1.75}~,
\end{eqnarray}
where $q\leq 1$ is the binary's mass ratio,  $M_{\rm bin}$ its mass,  $a_{\rm h}$
the semi-major axis when the binary first becomes ``hard'' (i.e. tightly bound), 
$a_{\rm gr}$ the separation at which gravitational-wave emission starts driving the binary's evolution
(cf. explicit expressions for $a_{\rm h}$ and $a_{\rm gr}$ in \citet{2013degn.book.....M}), $V_{\rm kick}$ 
 the post-merger kick velocity (computed using the fit to numerical-relativity simulations presented in \citet{2010ApJ...719.1427V}), 
and $V_{\rm esc}$ the escape velocity from the central parts of the galaxy (computed using the bulge and \NC\ density profiles).

 The first term in equation~(\ref{ejmass}) accounts for the mass deficit 
before the \MBH \ binary becomes hard~\citep{2006ApJ...648..976M}; 
 the second  is the mass ejected from $a_{\rm h}$ to $a_{\rm gr}$;
 and the third represents the mass deficit
 generated as the kicked \MBH \ heats up the surrounding core~\citep{2008ApJ...678..780G}.

We also account for the tidal disruption of \NC s by \MBH s
during galaxy mergers. More precisely, in a merger between a galaxy containing a \MBH\ and one
containing only a \NC, the \NC \ is  tidally disrupted before
reaching the center of the newly formed galaxy, if the \MBH \ mass is $\gtrsim 10^{8-9}~M_{\odot}$.
This is the same process described above for the infall of stellar clusters.

Next, we present our model's predictions, with the free parameter $f_{\rm gc}$ set to
its Milky Way value, i.e. $f_{\rm gc}\approx 0.07$~\citep{2012MNRAS.426.3008K}.
We have checked the robustness of our results 
against the MBH seed model (i.e. light vs heavy seeds, with several halo occupation numbers), 
a different value for $f_{\rm gc} \lesssim 0.2$, a variable
$f_{\rm gc}$~\citep[set to 0.07, 0.04 and 0.5 
in disk, quiescent and starburst galaxies respectively;][]{2012MNRAS.426.3008K}
and other details of our model (i.e. merger-tree resolution, initial redshift of the simulations, 
 AGN feedback prescriptions, etc.).
Indeed, we will show
that the crucial ingredient to reproduce the observed global to \NC \ correlations is the scouring effect due to \MBH\
binaries described by equation~(\ref{ejmass}).  More details of our model will be presented elsewhere~\citep{long_paper}.

\begin{figure}
\centering
\includegraphics[width=2.5in,angle=0.]{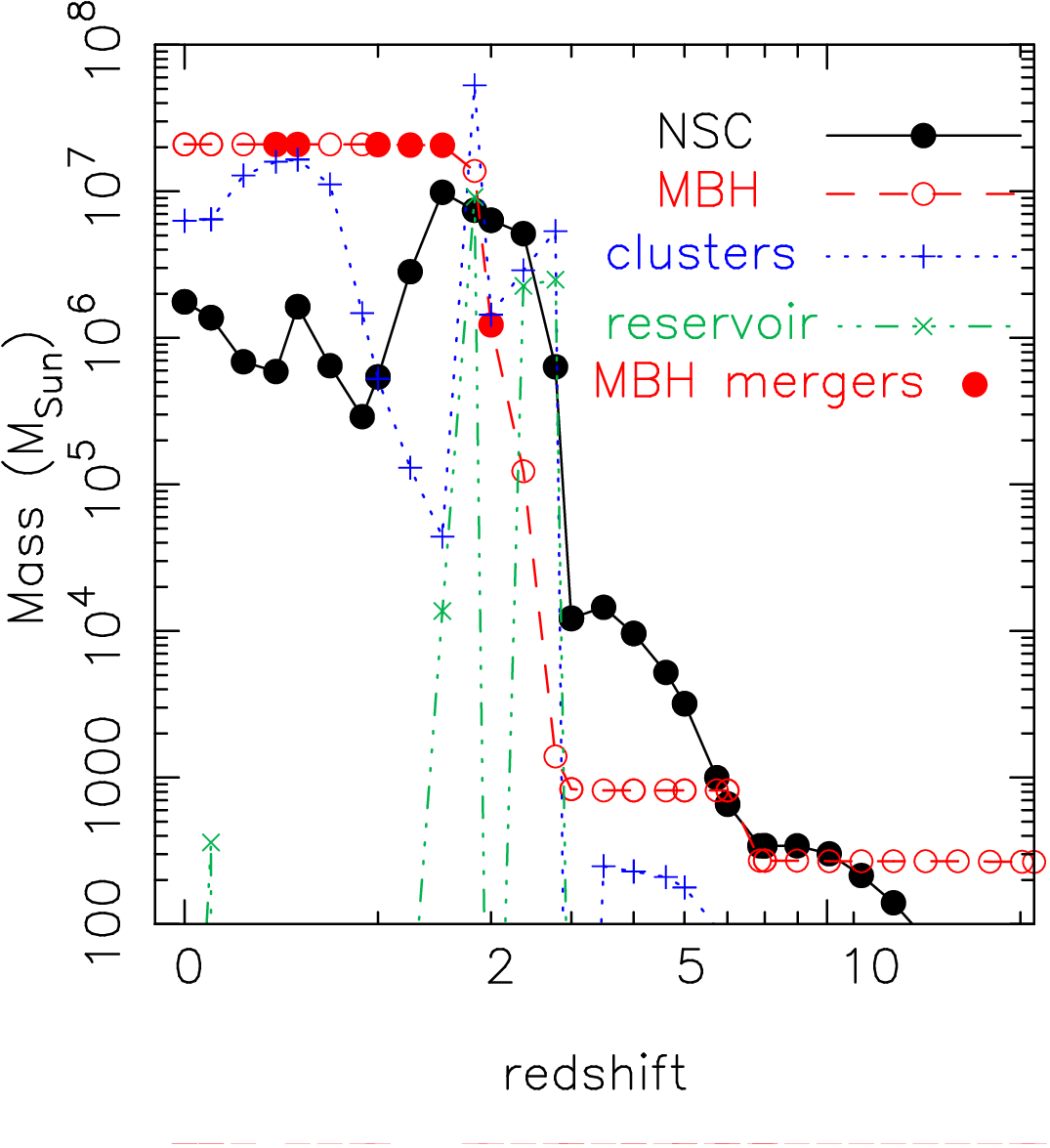} 
\caption{Main progenitor evolutionary track of   
  a galaxy  whose \NC\ is partially disrupted by  \MBH \ mergers. We show the mass of  the \NC ,
of the \MBH , of the central gas reservoir, and the total mass in stellar clusters, as a function of $z$.
\label{Z-ev1}}
\end{figure}

\begin{figure*}
\centering
\includegraphics[width=3.19in,angle=0.]{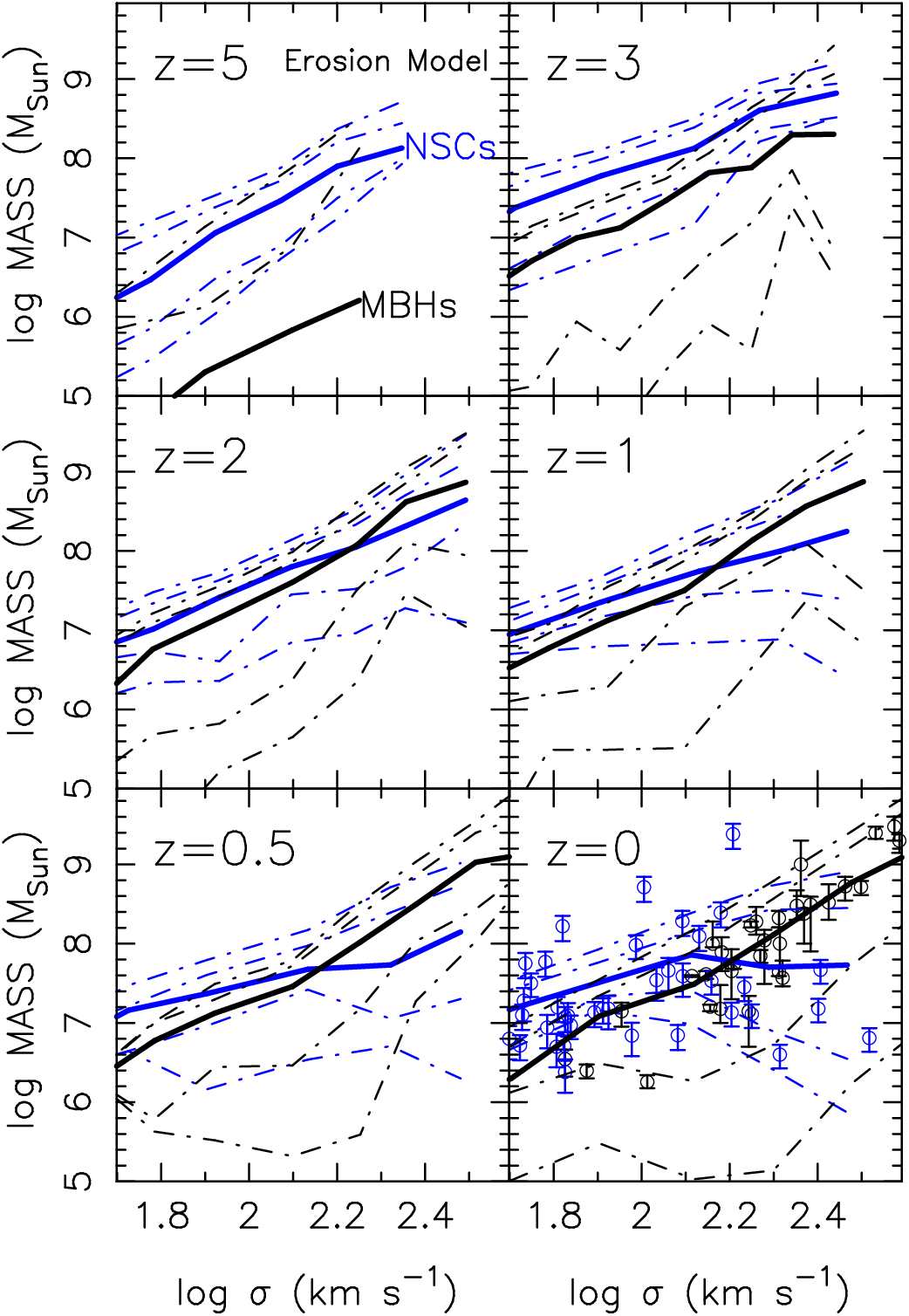} 
\includegraphics[width=2.8in,angle=0.]{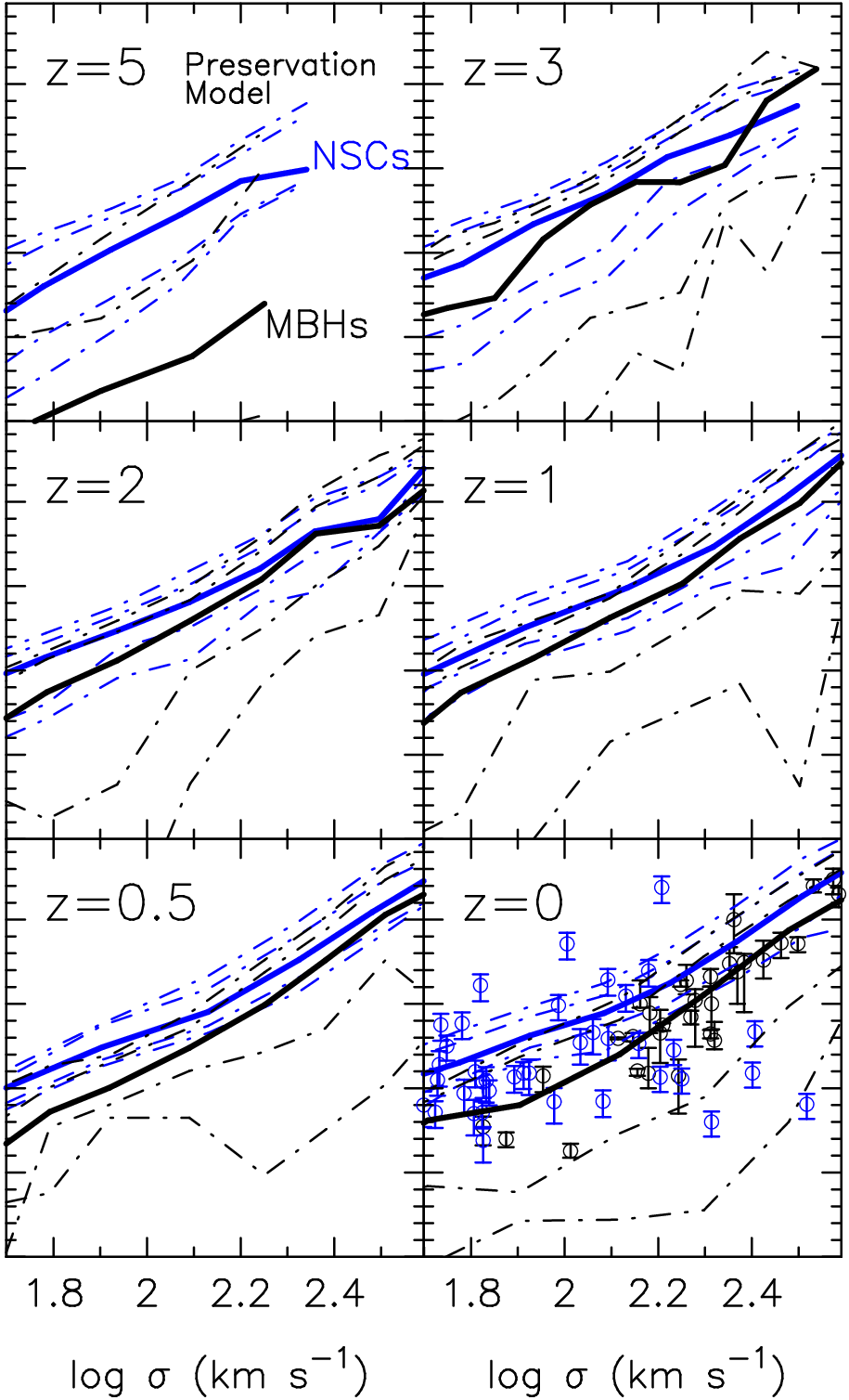}
\caption{
  Redshift evolution of \NC \ and \MBH \ $M-\sigma$ relations in our models
with (left panels) and without (right panels) mass ejection due to  \MBH \ binaries.
 The bottom right panels compare our predictions to data. Blue open circles are \NC s; black open circles are \MBH \ masses
from \citet{2002ApJ...574..740T}.  
Note the different evolution of the \NC\ mass vs $\sigma$ relation in the two models, at $z\lesssim2$.
Since reliable  \MBH \
 mass measurements have been obtained almost exclusively for main early type
galaxies~\citep[e.g.,][]{2005SSRv..116..523F},
in this analysis we have neglected \MBH s and \NC s in satellite galaxies, and only included galaxies with bulge to total mass ratio $B/T>0.7$, in order to compare with the observed 
\MBH \ $M-\sigma$ relation. 
However,  if all galaxies were included, the  \NC \ scaling correlations
would be very similar to those shown here (see Fig.~\ref{scaling-r},
which was produced by considering all galaxies).
\label{Z-ev2}}
\end{figure*}

\section{Results}
We describe the results from two realizations of our galaxy-formation model, 
namely our fiducial model, which accounts for the ``\NC \ erosion'' due to MBH binaries
through equation~(\ref{ejmass}); and a ``\NC \ preservation'' model where that effect is absent (i.e.
 we set $M_{\rm ej}=0$ throughout the entire cosmic history).

Our model's results are compared to a literature compilation of \NC\ masses.
 We constructed our sample of \NC\ objects by combining  data from Table 1 of
\citet{2013ApJ...763...76S}, Table  2 of \citet{Erwin:2012}, Table 2 of \citet{Neumayer:2012} 
and by estimating  \NC\  masses for the early type galaxies in the Fornax cluster catalog of   
 \citet{2012ApJS..203....5T}. The masses of the \NC s in  \citet{2012ApJS..203....5T} sample
 were calculated from the ($g-z$) colors given in that paper and from the relations of
 \citet{2003ApJS..149..289B}.  The 
 galaxy velocity dispersion and bulge mass were also obtained
from these papers or, when not available, from the Hyperleda database~\citep{2003A&A...412...45P}.

The upper panels of Figure~\ref{scaling-r} compare the observed 
$M_{\rm \NC}-\sigma$ and $M_{\rm \NC}-M_{\rm bulge}$ relations to  our ``\NC \ erosion'' model.
 We note the excellent  agreement of our predictions with observations. 
In particular, the predicted scaling relations appear to broaden at higher velocity dispersion, as
do the data.  
The broadening of the data appears to be caused by  the subsample
of galaxies with determined \MBH \ mass (red points in Figure~\ref{scaling-r}). The
nuclear to global correlations in these galaxies appear to flatten or even decline
for spheroids with higher velocity dispersion,
 suggesting that the \MBH s have played an important role in affecting the properties of their host \NC s
in these galaxies.

The blue lines in the middle panels of  Figure~\ref{scaling-r} shows the scaling relations obtained from
the ``\NC\ preservation'' model.
Note that the broadening at high $\sigma$'s present in the upper panels disappears. This demonstrates that the 
broadening of the scaling relations seen in the 
``\NC\ erosion'' model  is
a consequence of dynamical ejection of stars from  coalescing \MBH \ binaries.
Together with the model's predictions, we plot
$M_{\rm CMO}=M_{\rm \NC}+M_{\rm \MBH}$
vs $\sigma$ and $M_{\rm bulge}$,
 for nucleated galaxies with measured \MBH\ mass or an upper limit 
to it.  
\MBH\ masses and upper limits 
were taken from Table  1 of \citet{Erwin:2012}, table 2 of \citet{Neumayer:2012},
 and Table 1 of \citet{2009MNRAS.397.2148G}.
The red lines show fits to these data. 
As can be seen, when  the disruptive effect of \MBH \ binaries is not included,  our model
recovers the fitted functional form of the relation between $M_{\rm \CMO}$ and galaxy properties, 
both in slope and normalization. 

The fact that our ``\NC\ preservation'' model predicts 
scaling  correlations that are consistent with relations involving $M_{\rm \CMO}$
is not coincidental, but rather quite revealing. Indeed, the mass ejected by 
a \MBH \ binary  is of the  order of the mass of the binary itself,
with only a weak dependence on the binary's mass ratio and
the initial density distribution of stars~\citep{2006ApJ...648..890M}. 
Hence,  the mass ejected in
$N$ \MBH \ mergers is approximately $M_{\rm ej}\approx N M_{\rm \MBH} $,
with $M_{\rm \MBH}$ the final \MBH\ mass.

Because infall of gas to the nuclear region can rapidly rebuild a  
\NC, the relevant quantity to estimate
is the number of \MBH\ mergers since 
the era at which most of the gas was
depleted from the galaxy. \citet{2002MNRAS.336L..61H} 
compute this quantity using semi-analytic models for galaxy mergers 
similar to ours. They find that the number of  comparable-mass MBH mergers  is weakly dependent on
 galaxy luminosity, and has a small dispersion around a median of
$1$, with values   $\lesssim4$ even for the most luminous
galaxies.  
Accordingly, derived mass   deficits in ``core galaxies''  are also found to be peaked 
 around a value of $\approx M_{\rm \MBH}$~\citep{2004ApJ...613L..33G}.
The total amount of ejected stellar mass
is  therefore determined mainly by, and is roughly
equal to,  $M_{\rm MBH}$. Therefore,
$M_{\rm CMO}=M_{\rm \NC}+M_{\rm \MBH}$ can  be regarded  as an
approximate estimate  (with uncertainties of order a few) of
the \NC \ mass that was in place before mass erosion 
due to  \MBH \ binaries
became important. Our models do indeed confirm this.

These facts explain why the ``\NC\ preservation'' model 
predicts scaling  correlations 
that are consistent with relations involving $M_{\rm \CMO}$,
and provide further  support for the idea that the broadening/bending 
of the observed correlations between \NC \  and host-galaxy properties is a consequence 
 of the nuclear mass erosion caused by MBH binaries.
Finally, the lower panels of Figure~\ref{scaling-r}  show  the
relations between $M_{\rm \CMO}$
and galaxy properties  in the \NC \ erosion model,
for nucleated galaxies with a \MBH .  
Clearly, the choice of plotting the total  $M_{\rm \CMO}$ mass
significantly reduces the broadening  that characterized the $M_{\rm \NC}$ relations (upper panels),
but the scatter is slightly larger than in the ``\NC\ preservation'' model (middle panels).
This is  expected  since $M_{\rm ej}$ only equals
the final \MBH\ mass to first approximation (see equation~[1]).

The ancestry of a \NC \  includes complicated processes leading to its growth, disruption, and
regeneration if new material is accreted after a disruptive merger.
In Figure~\ref{Z-ev1}  we illustrate  the typical evolution of \NC s as predicted by our fiducial ``\NC \ erosion''  model.
Since the effect of \MBH \ binaries and mergers on \NC's is most important for large galaxies (where they cause the broadening/bending
of the correlations between \NC \ and host-galaxy properties), we choose a galaxy 
with spheroid mass $\sim 2\times 10^{10} M_{\odot}$,
at which the \NC \ erosion becomes significant.
In this example, the \MBH \ grows by
short-lived accretion events triggered by bar instabilities of
the host's galactic disk (at $z\sim6-7$) or by major galactic mergers (at $z\sim 1-3$; note the 
corresponding growth of the reservoir following the merger-driven infall of cold
gas to the nuclear regions). The \NC \ grows more gradually through a combination
 of in-situ star formation (especially at $z\sim 1-3$, when the growth of the reservoir triggers nuclear star formation),
and infall of stellar clusters. The latter channel dominates over the former at high redshift, and is also enhanced at $z\sim 1-3$, since major galaxy mergers also
cause violent bursts of star formation, which in turn
enhance stellar-cluster formation. Finally, at $z\lesssim 1-2$ the \NC \ gets eroded by a series of minor \MBH \ mergers,
but re-forms thanks to a steady infall of stellar clusters, which form as a result of quiescent star-formation activity as the galactic disk re-grows. 

Figure~\ref{Z-ev2} shows the redshift evolution of the $M-\sigma$ relations for both \MBH s and \NC s,
in the  ``NSC erosion'' model (left panels), and in the ``NSC preservation'' model (right panels).
In both models, the \MBH \ growth appears inefficient at high redshifts relative
to \NC \ growth. At redshift $z\lesssim 2$, after the epoch of bright quasars,
 \MBH s become the  dominant nuclear component in the most massive galaxies. 
Therefore,  it is only at relatively late cosmic epochs that \MBH \ mergers
become efficient at scouring their host clusters, in the ``\NC \ erosion'' model.
As can be seen, the ``NSC preservation'' model clearly over-predicts 
the mass of \NC s in the brightest nucleated spheroids, while  the ``\NC \ erosion'' 
model  at $z=0$ is in good agreement with  the local $M-\sigma$ relation for
both \MBH s and \NC s.

\section{Conclusions}

A comparison  of our  results  to observational data reveals that
 the \NC \ scaling correlations in the local Universe
 carry an imprint of the merger and growth  history of their companion   \MBH s.
\NC s are significantly eroded by \MBH \ binaries,
 causing a broadening of the observed \NC \ empirical correlations in high-mass  galaxies and at low redshifts.
 
The \NC \ scaling correlations
can potentially  be used to probe different \NC \ evolutionary 
models, and also to place constraints
 on the merger and growth history of their host galaxies.
However, the slope of the observed correlations involving  \NC \ masses is
 lowered by the inclusion, at the high-$\sigma$ end,
of \NC s that were partly eroded by \MBH s. 
Hence, such relations cannot be used to put
reliable constraints on different \NC \ formation models
without also taking into account  the scouring effect of \MBH \ binaries. 

We argue that
the stellar mass removed from the center during mergers 
is of order of $M_{\MBH}$, and indeed
we find that by replacing  $M_{\rm \NC}$ with $M_{\rm \CMO}=M_{\rm \NC}+M_{\MBH}$,
the observed scaling correlations  agree remarkably well with those predicted by 
models that do not account for the
scouring effect of \MBH \ binaries.
Our findings   provide a natural explanation to why relations between
$M_{\rm \CMO}$ and galaxy properties for nucleated galaxies show 
significantly less scatter than 
relations involving $M_{\rm \NC}$ alone.
The relations between $M_{\rm \CMO}$ and galaxy properties 
are much shallower ($\sim \sigma^2$) than the same correlations 
for \MBH s~($\sim \sigma^5$). 
Our results also explain the well known  transition from  \MBH  --  to \NC  --
dominated galaxies  as one proceeds from dwarfs to giant ellipticals,
without the need of invoking competitive feedback processes from 
 young \NC's \ and/or AGN activity. Therefore, the formation of \MBH s and \NC s appear \textit{not} to be
mutually exclusive, with \NC s being  modified after their formation by the merger evolution of their companion \MBH s.

\acknowledgments
We acknowledge support from a
CIERA  fellowship at Northwestern University
 (to F.A.); from the European Union's Seventh Framework Programme (FP7/PEOPLE-2011-CIG)
through the  Marie Curie Career Integration Grant GALFORMBHS PCIG11-GA-2012-321608 (to E.B.); 
from   ERC project 267117 (DARK) hosted by Universit\'e Pierre et Marie Curie -- Paris 6 and at 
JHU by National Science Foundation grant OIA-1124403  (to J.S.). Computations 
were performed on SciNet, and on the Horizon Cluster at the IAP.

  \end{document}